**Photoconductive Effects in Single Crystals of BaZrS₃**

*Boyang Zhao, Huandong Chen, Ragib Ahsan, Fei Hou, Eric R Hoglund, Shantanu Singh, Huan Zhao, Han Htoon, Andrey Krayev, Maruda Shanmugasundaram, Patrick E Hopkins, Jan Seidel, Rehan Kapadia, and Jayakanth Ravichandran*

Boyang Zhao, Huandong Chen, Shantanu Singh, Jayakanth Ravichandran
Mork Family Department of Chemical Engineering and Materials Science, University of Southern California, Los Angeles, California, 90089, USA
E-mail: j.ravichandran@usc.edu

Ragib Ahsan, Rehan Kapadia, J. Ravichandran
Ming Hsieh Department of Electrical Engineering, University of Southern California, Los Angeles, California, 90089, USA

Fei Hou, Jan Seidel
School of Materials Science and Engineering, University of New South Wales, Sydney, NSW, Australia

Eric R Hoglund, Patrick E Hopkins
Department of Mechanical and Aerospace Engineering, University of Virginia, Charlottesville, Virginia, 22807, USA
Department of Materials Science and Engineering, University of Virginia, Charlottesville, Virginia, 22904, USA

Huan Zhao, Han Htoon
Center for Integrated Nanotechnologies (CINT), Materials Physics and Applications Division, Los Alamos National Laboratory, Carlsbad, New Mexico, 88220, USA

Andrey Krayev,
HORIBA Instruments Inc., Novato CA, 94949, USA

Maruda Shanmugasundaram
HORIBA Instruments Inc., Piscataway NJ, 08854, USA

Eric R Hoglund

Center for Nanophase Materials Sciences, Oak Ridge National Laboratory, Oak Ridge, Tennessee 37830, USA

Patrick E Hopkins

Department of Physics, University of Virginia, Charlottesville, Virginia, 22904, USA

Jayakanth Ravichandran

Core Center of Excellence in Nano Imaging, University of Southern California, Los Angeles, California, 90089, USA

**Abstract**

Chalcogenide perovskites, such as $BaZrS_3$, are emerging semiconductors with potential for high photovoltaic power conversion efficiency. The role of defects in the efficiency of the generation and collection of photo-excited carriers has not been experimentally investigated extensively. We study the effect of processing-induced defects on the photoconductive properties of single crystals of $BaZrS_3$. We achieved ohmic contacts to single crystals of $BaZrS_3$ and observed positive surface photovoltage, which is typically observed in *p*-type semiconductors. However, mechanical polishing of $BaZrS_3$ to remove the surface oxide leads to dense deformation grain boundaries and leads to trap-dominated photoconductive response. In comparison, ohmic contacts achieved in cleaved crystals leave fewer deformation defects and greatly improve opto-electronic properties. Defect-controlled crystal growth and contact fabrication are potentially limiting factors for achieving high photon-to-excited electron conversion efficiency in $BaZrS_3$.

## 1. Introduction

From silicon,[1] GaAs,[2] to halide perovskites,[3,4] semiconductors with excellent optoelectronic properties have laid the foundation of modern photonics. Recently, a new class of inorganic semiconductors, chalcogenide perovskites,[5–8] has drawn significant attention due the characteristics such as earth-abundant and non-toxic composition,[9] tunable bandgap in the visible and infrared energies,[5] and thermal stability.[10] Polycrystalline[11–14] and epitaxial[15,16] thin films of BaZrS$_3$ have been used to realize prototypical metal-semiconductor heterojunctions with appreciable photoresponsivity. Solution-processed BaZrS$_3$ thin films[18–20] have been explored to realize low-cost optoelectronic devices. However, the role of extended defects *i.e.*, grain boundaries, dislocations, and crystalline domains and interfaces during thin film growth and device fabrication has not been systematically studied yet. Hence, the intrinsic optoelectronic performance limits have not been studied, despite the availability of single crystals of BaZrS$_3$.[20,21]

In this work, we use single crystals of BaZrS$_3$ grown with BaCl$_2$ flux to establish optoelectronic device fabrication protocols. The electrical contacts to bulk BaZrS$_3$ crystals were fabricated after removing the oxide layers using mechanical polishing or cleavage. We studied the work function of BaZrS$_3$ by Kelvin Probe Microscopy (KPFM), which shows an obvious decreasing trend under optical excitation, resembling a *p*-type semiconductor. Conversely, two terminal ohmic electrical contacts to polished BaZrS$_3$ crystal show trap-dominated photo-response in time-resolved, power-dependent, or spatially mapped photocurrent measurements. The deep traps are presumably related to extended defects caused by mechanical deformation while polishing the surfaces. Cleaving the crystals to expose fresh surfaces to electrical contacts showed significant improvement in the optoelectronic properties, in agreement with this notion. These results show the limits posed by extended defects during materials processing for optoelectronic devices of BaZrS$_3$ and show the need to develop processing techniques to access the intrinsic performance limits of chalcogenide perovskite optoelectronic devices.

## 2. Results and Discussion

### 2.1. Crystal Growth and Opto-electronic Properties

Single crystals of BaZrS$_3$ were grown in an evacuated sealed ampoule using BaCl$_2$ flux, as reported earlier[21,22] but with slightly lower cooling rate (Methods I). We were able to achieve relatively larger single crystals of BaZrS$_3$ using the slower cooling rate (*i.e.,* cubes with sides of ~250 μm). First, we characterize the opto-electronic quality of the crystals using time-resolved photoluminescence (TRPL). The bi-exponential relaxation time of photo-excited states, as measured by TRPL for BaZrS$_3$ crystal [**Figure S1**] was $\tau_1 = 0.675 \pm 0.005$ ns, $\tau_2 = 3.97 \pm 0.04$ ns, shorter but comparable to the SRH recombination lifetimes (SRH) of ~50 ns [21] ($\tau_1 = 1.069 \pm 0.009$ ns, $\tau_2 = 8.00 \pm 0.15$ ns) measured on earlier BaZrS$_3$ crystals. Considering the similarity of crystal size and surface conditions, defects randomly introduced during the cooling cycle could have caused subtle differences in the TRPL response.

Next, we attempted to fabricate electrical contacts to single crystals of BaZrS$_3$ to characterize their opto-electronic properties. As grown crystals, after extraction from the flux, invariably have a thin insulating layer of sulfate, sulfite, or oxide on the surface. Presumably, this layer is formed during exposure to water and/or air, while extracting the crystals from the salt flux, and severely limits our ability to make ohmic contacts. To overcome this issue, we resorted to mechanically polishing off (optimized by crystal embedding, details see Methods) the insulating layer [**Figure 1a**] or cleaving it [inset of **Figure 4a**] right before depositing or transferring metal contacts. Both approaches are discussed in detail in Methods, and both achieve ohmic contacts or nearly ohmic contacts to bulk BaZrS$_3$.

First, we leveraged stable electrical contacts to perform static surface potential measurements on BaZrS$_3$ using the Kelvin Probe Microscopy (KPFM)[22]. We will use optical illumination during the KPFM studies to track the evolution of the surface potential to learn about the accumulation of the photo-excited carriers. BaZrS$_3$ crystals were electrically connected to AFM disks with silver paint on the backside. Contact potential difference (*CPD*, illustrated in **Figure 1b**) of BaZrS$_3$ is then measured [**Figure S2**] on the top surface [**Figure 1c**, **Figure 1d**] by KPFM. *CPD* map of BaZrS$_3$ and the highly oriented pyrolytic graphite (HOPG)[23], were measured under the same condition to calibrate the work function values determined from these measurements. BaZrS$_3$ work function is thus extracted to be $4.44 \pm 0.02$ eV (Supporting Information Section I). This is a slightly higher work function than Ti (4.33 eV) and Ga (4.32 eV)[24]. Both metal contacts were tested and established as ohmic contacts to BaZrS$_3$.

Next, we introduce optical illumination in the same region of KPFM to investigate the surface photovoltage (SPV) accumulation on the surface of an as-grown (**Figure S2**, SPV up to +0.18 V) and a polished (**Figure 1a**, SPV up to +0.15 V) BaZrS$_3$ surface. As the charge neutrality level of semiconductor surfaces bends the conduction and valence band near the free surface depending on the type of carrier in the semiconductor [**Figure 1b**], photo-excited carriers accumulated near the surface will dictate the band curvature, which typically leads to a positive $\Delta CPD$ in the case of *p*-type semiconductors or negative $\Delta CPD$ in the case of *n*-type semiconductors[25]. In the case of BaZrS$_3$ [**Figure S3**], 785 nm excitation leads to lower $\Delta CPD$ compared to 532 nm or 638 nm excitation, as 785 nm has energy below the bandgap of BaZrS$_3$ ~ 1.7-1.95 eV[7,16,26,27], but a consistent positive SPV (decreasing work function) is observed, under all three excitation wavelengths. Further, the increase in SPV scales with the increase in the excitation power, which saturates beyond a certain wavelength-dependent threshold. [**Figure 1e**, **Figure 1f**]. We thus conclude that the surface photovoltage characteristics of BaZrS$_3$ are consistent with a *p*-type semiconductor.

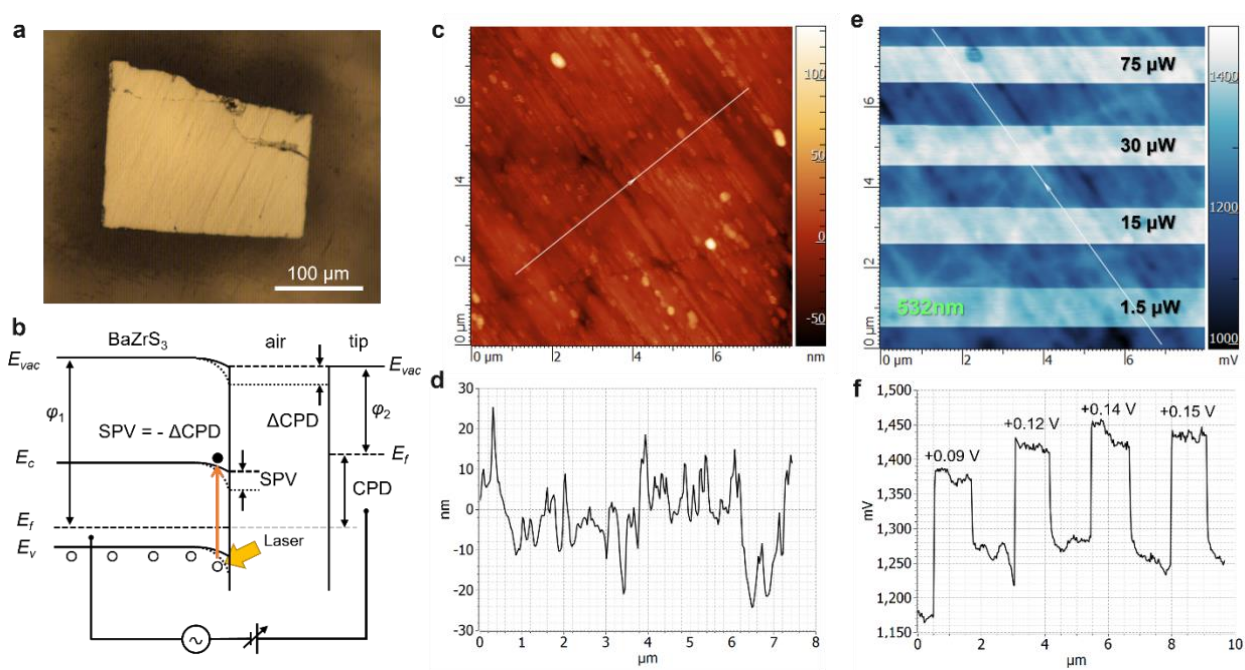

**Figure 1**. Surface photovoltage (SPV) characterization of BaZrS$_3$ *via* KPFM. (a) Optical image of the polished top surface of BaZrS$_3$. (b) Illustration of the band diagram of a *p*-type semiconductor surface in electrical contact with the metallic tip of the KPFM with and without illumination. (c) AFM topography of polished BaZrS$_3$. The surface roughness $R_a$ was lower than 7 nm. (d) The spatial dependence of the topography of the polished sample corresponds to the white line noted in (c). (e) *CPD* map of the surface of BaZrS$_3$ crystal across the same area in (c). The optical excitation power is noted, and dark regions correspond to dark conditions without optical illumination. The *CPD* scales with optical excitation power. (f) The spatial dependence of *CPD* along the white line noted in (e).

## 2.2. Photoconductivity studies on polished crystals

Next, we fabricated two-terminal devices to study the photoconductivity response in BaZrS$_3$ single crystals. We made Ti/Au (details see Methods), which is nearly ohmic, and a typical device is shown in the inset of **Figure 2a**. We observed relatively linear DC current-voltage, or I-V characteristics under dark conditions, and a typical I-V curve is shown in **Figure 2a**. We observed positive photocurrent ($I - I_{dark}$, hereafter noted as photocurrent for brevity) under the illumination of 405 nm, 532 nm, and 633 nm wavelengths [**Figure 3b**], and a sharp drop in photocurrent was observed [inset of **Figure 3c**] when we used photon energies (of 710 nm and 785 nm) below the bandgap of BaZrS$_3$[7]. We normalized the power density in terms of sun power (1 sun = power of AM 1.5 illumination ~ 1000 W/m$^2$) [28] to compare the responsivity between different wavelengths of light used here. We observed a responsivity of up to ~0.3 A/W at 10 V under 532 nm excitation, equivalent to a ~0.18 external quantum efficiency (EQE). This is possibly led by either an efficient minority carrier photoexcitation or accumulated long-lifetime defect traps like BaZrS$_3$ thin films[16].

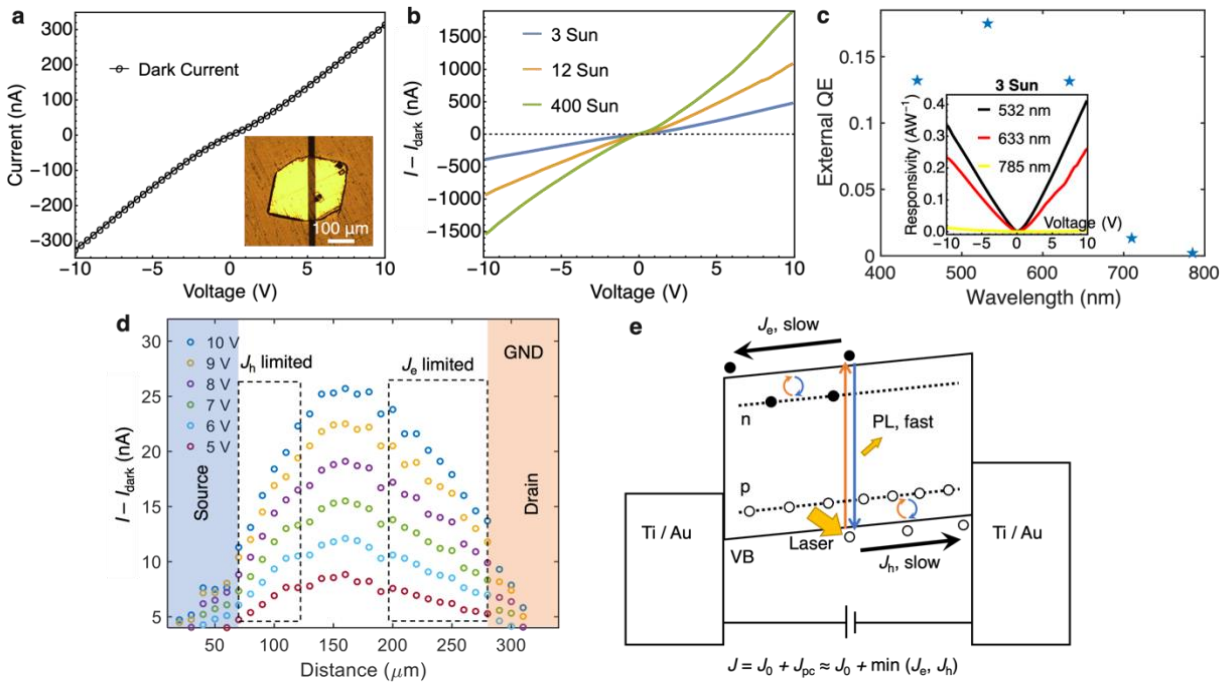

**Figure 2**. Photoconductivity studies on polished BaZrS$_3$ crystals. (a) I-V characteristics under dark conditions. Inset is the optical image of a device with two-terminal contacts to a BaZrS$_3$ crystal used in the measurements. It shows relatively linear I-V characteristics. (b) Measured enhancement in the current upon optical illumination in the active region of the device between the contacts. Optical illumination leads to an increase in photocurrent with increasing optical power density. (c) Wavelength dependence of responsivity or external quantum efficiency (QE)

derived from the photocurrent values measured as shown in (b). (d) Spatial dependence of photocurrent across the active region. The maximum photocurrent value occurs when the illumination is at the center between the two contacts, irrespective of the applied bias between the contacts. This suggests $BaZrS_3$ not to be degenerately doped in nature. The trap-dominated photo response is illustrated in (e) using a schematic of the band diagram. Trapped defects occupied by photo-excited carriers give rise to additional long-lived carriers than carriers excited to the band edges in $BaZrS_3$.

### 2.3. Scanning Photocurrent Microscopy Studies

Next, we performed a scanning photocurrent microscopy (SPCM) study by scanning the focused (~ 0.8 μm diameter spot size, details in Supplementary Section IV) 532 nm laser excitation across the active channel from source to drain while recording the photocurrent response to learn about the nature of the photoexcited carriers. **Figure 2d** shows the SPCM mapping results under different DC bias voltages. The maximum photocurrent was observed in the middle and roughly equidistant from the source and drain, irrespective of the applied voltage bias, which is inconsistent with the SPCM behavior expected from *p*-type nor *n*-type semiconductors[29] (*n*-type Silicon verified in Supplementary Section IV). For degenerately doped semiconductors, we expected the maximum photocurrent to occur near the source (*p*-type) or drain (*n*-type) electrodes under a $+V_{source}$ ($V_{drain} = 0$), and such photocurrent profile is generally inversed under a $-V_{source}$. Past studies showed that the photo-excited minority carrier mean-free path (~5 μm[21]) for similar single crystals of $BaZrS_3$ is much smaller than the channel length (~ 200 μm). If one were to estimate the photo-excited carrier mean free path using the shallow decay in the spatial dependence of photocurrent, the SPCM-derived mean free path would be inconsistent with the much shorter mean free path observed in TRPL. Hence, we presume that a long lifetime is consistent with the presence of a large concentration of deep traps. Such traps have a much longer lifetime than photo-excited carriers and thus lead to long relaxation time, as illustrated in **Figure 2e**. The total steady-state photocurrent is then dominated by the trapped carriers that eventually drift to opposite electrodes before getting collected.

### 2.4. Electron Microscopy Studies

To understand the origin of the discrepancy in the deduced mean free path, and whether any extended defects were created during the processing of the samples for the photoconductance studies, we carried out scanning tunneling electron microscopy (STEM) studies near the surface of a polished BaZrS$_3$ [**Figure 3a**] and a cleaved BaZrS$_3$ [**Figure 3b**]. **Figure 3a** shows a large number of planar defects right beneath the polished surface, which from selected area diffraction (Supporting Information Section V) are determined to be domain walls. Considering the absorption coefficient of BaZrS$_3$ for 532 nm is ~2×10$^6$ 1/cm, photo-excited carriers are largely generated within a penetration depth of ~50 nm. Thus, extended defects observed here down to 300 nm would dominate the photoconductance response of the polished crystals of BaZrS$_3$. In comparison, **Figure 3b**, STEM near a cleavage surface of BaZrS$_3$, does not feature any noticeable cracks or other extended defects near the surface and the projection of atomic planes can be resolved within a few nm from the surface. Thus, less defective cleaved BaZrS$_3$ surfaces are expected to have faster photocurrent response with lower number trap states. Nevertheless, we may also note that orthorhombic[20] BaZrS$_3$ does not have an easy-to-cleave crystalline plane. The presumed (010) cleavage surface will be uneven during cleavage, making it non-ideal for avoiding extended defects. Thus, we anticipate some concentration of defects that would lead to traps either in the bulk or at the surface.

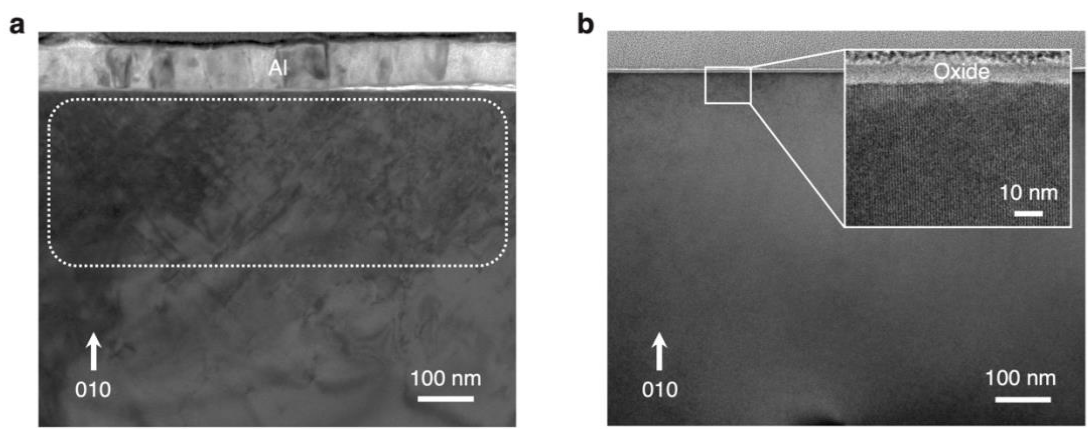

**Figure 3**. Extended defects introduced by mechanical polishing BaZrS$_3$. (a) STEM image of the polished BaZrS$_3$ covered by Al. Extended defects such as dislocations, stacking faults, and grain boundaries are populated down to ~300 nm into the surface, marked by a dotted box. (b) STEM image of the cleaved BaZrS$_3$ without any capping layer. No cracks or extended deformation are observed across the cleaved surface. Inset observed uniform BaZrS$_3$ (101) or (10$\bar{1}$) atomic planes within a few nm beneath the oxide surface.

## 2.5. Comparison of Cleaved and Polished Crystals

Next, we compare the photoconductance response of the polished and cleaved crystals by performing photoconductance studies on cleaved BaZrS$_3$ crystals. First, the cleaved crystals of BaZrS₃ show much lower dark conductance [**Figure 4a**] but significantly better linearity in I-V characteristics compared to the polished crystals of BaZrS₃ under both dark and illuminated conditions. Moreover, **Figure 4b** shows the time-resolved photocurrent measurements on polished and cleaved BaZrS₃ crystals, where clear differences in the time constant in the rise and decay of the photocurrent. It is evident that avoiding extended defects formed during the processing of BaZrS₃ can dramatically improve the rise/decay time of the photocurrent from 18/26 s to 2/2 s; the latter is comparable to the epitaxial BaZrS₃[16]. Further, the contribution of the trap-contributed photo-conductance can be deduced from the power law relationship between the induced photocurrent and incident power of light, noted as $I - I_{dark} \propto P^{\theta}$, where $\theta = 1$ is the ideal photocurrent current generation and smaller $\theta$ reflects more trap-domination photocurrent generation. **Figure 4c** shows the relationship between photocurrent and excitation power $P$ and the fitted powder law dependence of polished and cleaved crystals of, and epitaxial thin films of [16] BaZrS₃. It is worth noting that device geometries here are neither exactly comparable nor optimized. The power law scaling for cleaved BaZrS₃ crystals reaches $\theta = 0.67$, which is comparable to $\theta = 0.71$ of epitaxial BaZrS₃[16], which is significantly higher than $\theta = 0.29$, observed for polished crystals of BaZrS₃. Thus, we show that mechanical deformation-induced extended defects play a significantly detrimental role in the optoelectronic properties of BaZrS₃.

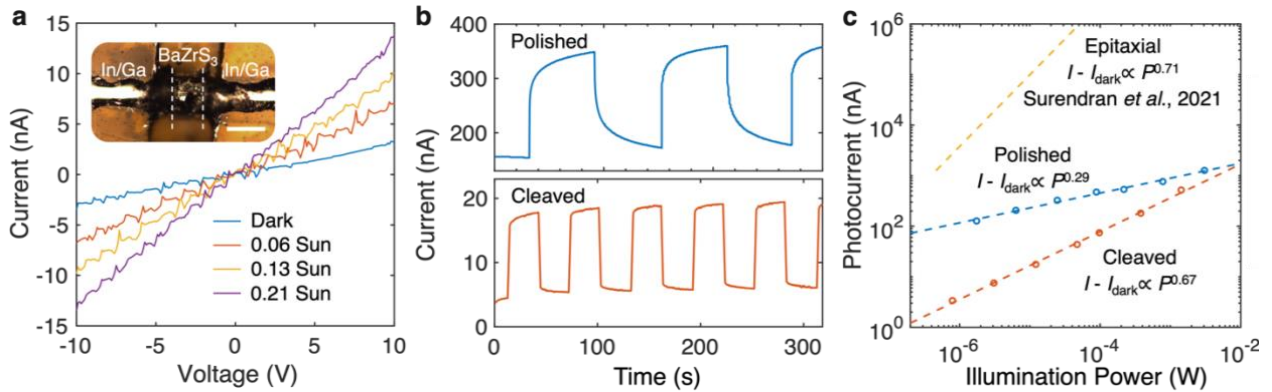

**Figure 4**. Photoconductance response of polished and cleaved crystals of BaZrS₃. (a) Dark and illuminated I-V characteristics of cleaved crystals of BaZrS₃. Halogen incandescent lamp with measured 0.06, 0.13, and 0.21 sun power densities were used for the measurements. Inset is the picture of two terminal junctions to a cleaved BaZrS₃, scale bar 200 μm. (b) Transient photocurrent response to 0.24 sun power white light from halogen lamp. Rise time (time to reach 90% photocurrent from dark current after illumination on): $\tau_{polished} = 18$ s, $\tau_{cleaved} = 2$ s; Decay time (time to reach 10% of the photocurrent after switching off excitation): $\tau_{polished} = 26$

s, $\tau_{cleaved}$ = 2 s. Cleaved samples have a much faster photocurrent response. (c) Power law dependency of photocurrent with intensity of 532 nm illumination at +10V, fitted as $I$ - $I_{dark} \propto P^{\theta}$. Polished BaZrS$_3$, $\theta$ = 0.29, cleaved BaZrS$_3$, $\theta$ = 0.67, in comparison with epitaxial BaZrS$_3$ (Reproduced with permission.[16] Copyright © 2021 American Chemical Society) with 10 µm channel finger geometry electrodes.

## 3. Conclusion

We experimentally show that extended defects introduced by mechanical polishing are the major cause of the defect-dominated photoconductance behavior in single-crystal BaZrS$_3$. By avoiding such traps in cleaved BaZrS$_3$ crystals, the efficiency of separating photogenerated electrons and holes under DC voltage is greatly improved. Considering the relatively weak radiative recombination[7] and photoexcited carrier mean-free-path[21] well over the penetration length in single crystal BaZrS$_3$, electrode patterning, surface passivation, electron/hole heterojunction, and defect control would further improve the opto-electronic performance of single crystals of BaZrS$_3$. Optimization of the contacts, surfaces, and device geometry is the next critical step for achieving high-efficiency optoelectronic and photovoltaic devices based on BaZrS$_3$ and other chalcogenide perovskites.

## Methods

*Single Crystal Growth*: BaZrS$_3$ crystals were grown with BaCl$_2$ flux. 1g of Barium Chloride powder (Alfa Aesar, 99.998%) was mixed and ground along with 0.5g stoichiometric quantities of precursor powders, including barium sulfide powders (Sigma-Aldrich 99.9%), zirconium powders (STREM, 99.5%) and sulfur pieces (Alfa Aesar 99.999%) in a nitrogen-filled glovebox. They were then loaded into a quartz tube capped with ultra-torr fittings and a quarter-turn plug valve inside to avoid exposure to the air. The tube was then evacuated to around 10 mbar and sealed.

Precursors sealed in quartz tubes were heated to 1050 ℃ at a ramping rate of 100 ℃/h, held at 1050 ℃ for 100h, and then cooled to 850 ℃ at a cooling rate of 2 ℃/h, and allowed to naturally cool down by shutting off the furnace. The obtained samples were washed with cold deionized water repeatedly for 10 min to remove the excess flux and quickly dried in desiccators. The collected crystals are then sieved into different sizes and picked out from insoluble impurities.

*Time-resolved photoluminescence (TRPL)*: TRPL measurements were carried out with a HydraHarp400 time-correlated single photon counting system. PL was confocally excited by a 60 ps, 405 nm laser pulse at a 5 MHz repetition rate through a 50×objective with a numerical aperture of 0.7. The excitation power used was~100μW. The collected PL was spectrally filtered using a combination of a 630 nm long-pass filter and a 770 nm short-pass filter and detected by a pair of fast avalanche photodiodes with 16 ps time resolution.

*Kelvin Probe Microscopy (KPFM)*: KPFM measurements were carried out using an AFM (AIST-NT SmartSPM 1000) in the air. $BaZrS_3$ crystals cleaved or polished on the back side are connected to AFM metal specimen discs with silver epoxy. A nitrogen gun was used to remove particles on the as-grown top surface before the measurements. KPFM measurements were performed using a gold-coated Si tip with a 6 nm radius of curvature (HYDRA6R-100NG-10, APPNANO) as the probe in a noncontact mode with an AC voltage of −1 to +1 V. Crystals are illuminated under 450 nm, 532 nm, and 600 nm lasers to verify the SPV consistency.

KPFM under light illumination was carried out using XploRA Nano AFM-Raman Instrument from HORIBA Scientific. 532 nm, 638nm, and 785 nm lasers were aligned separately with the tip apex before KPFM. The XYZ coordinates of the objective were then changed by the software along with the laser selection itself in the middle of KPFM scans of polished $BaZrS_3$ crystals. ND filters were changed with increments for power dependence. SPV is extracted by the $\Delta CPD$ with and without turning on the laser.

*Electrical contacts to Polished Crystals*: The BZS crystals were first embedded in a polymeric medium to planarize the top surface such that regular mechanical polishing and photolithography processes can be applied. The crystal planarization processing is modified from the literature[30] as such: BZS crystals with a thickness of ~ 100 μm or below were picked up individually by a polydimethylsiloxane (PDMS) stamp. The crystal was then embedded in a UV-curable epoxy (NOA 61, Norland Products, Inc) medium on top of a sapphire substrate (5×5 mm) by attaching the PDMS stamp adjacent to the substrate, followed by a UV curing step (*i*-line UV lamp B-100, UVP for 30 min) to solidify the epoxy. The crystal was then gently polished with a 12000-mesh polishing cloth to remove the surface oxides. Electrodes of Ti/Au (20/300 nm) were then formed using standard photolithography and e-beam evaporation without much delay.

*Electrical Connections to Cleaved Crystals*: BaZrS$_3$ crystals were cut into half with a razor blade and loaded on Kapton double-sided tape with the cleaved surface facing up. The cleavage surface does not propagate along a specific crystalline orientation, so an uneven cleavage surface is commonly formed. Systematic electrode deposition is thus inaccessible. We then manually painted In/Ga eutectic to cover both ends of the cleaved surface to form two terminal ohmic electrical connections.

*I-V Characteristics and Optoelectronic Response*: I−V characterization and transient photocurrent I-t measurements were conducted using a semiconductor parameter analyzer (Agilent 4156C). Scans are carried out between DC -10V and +10V to avoid breakdown. Ambient steady-state and time-resolved photocurrent used Bausch & Lomb 20-watt halogen microscope bulb provided and controlled by Micromanipulator Probe Station to switch on or off the ambient illumination. The higher power densities were provided by solid-state lasers and attenuated by numeric apertures to control power intensity before being guided and focused on the active region of BaZrS$_3$.

Dark current (measured before excitation was introduced) was subtracted from light current. This gives us the unsaturated photocurrent, $I_{ph} = I - I_{dark}$. Incident optical power was estimated by:

$$P_{active} = A_{active} \times P_{spot} / A_{spot}. \tag{1}$$

Then responsivity and external quantum efficiency (EQE) at +10V DC are deducted:

$$R_\lambda = I_{ph} / P_{active}, \eta_{EQE} = R_\lambda \times E_{photon} \tag{2}$$

Wavelength dependence and quantitative photocurrent are achieved with 532 nm, 633 nm, and 785 nm static laser illumination provided and configured by a Renishaw Raman spectroscopy equipped with inVia confocal Raman microscope and diode lasers. 450 nm and 700 nm were performed using the Squidstat potentiostat.

*Scanning Photocurrent Microscopy (SPCM)*: SPCM was carried out within the same Renishaw Raman spectroscopy, but the laser spot size focused to ∼ 1 μm in diameter. The scanning is achieved by a motorized stage with an accuracy of 0.1 μm across the BaZrS$_3$ terminal junction at a step of 5 μm. The finer step size is tested near the electrodes to verify consistency. Two terminal I-V characteristic is measured with a Keithley 2400 source meter under dark or illuminated conditions.

*Scanning Tunneling Electron Microscopy (STEM)*: FIB samples were made using a Thermo Fisher Scientific Helios Dual Beam focused ion-beam. TEM samples were made from $BaZrS_3$ coated in Al to mitigate any charging or damage from the electron or gallium beams. Large regions of carbon and platinum were deposited with the electron beam prior to gallium beam exposure. Initial milling and cleaning steps were performed at 30 kV, which was sequentially decreased until the finishing energy of 2 kV. Aberration-corrected STEM imaging and Conventional TEM were performed on a Thermo Fisher Scientific Themis-Z STEM operating at 200 kV.

**Supporting Information**

Supporting Information is available from the Wiley Online Library or from the author.

**Acknowledgements**

We thank Prof. Rafael Jaramillo for useful discussions. This work in part was supported by the Army Research Office (ARO) under award numbers W911NF-19-1-0137 and W911NF-21-1-0327, and by the US National Science Foundation with award numbers DMR-2122070 and DMR-2122071. P.E.H. and E.R.H. appreciate the support from the Office of Naval Research, Grant number N00014-23-1-2630. R.A. acknowledges USC Provost Graduate Fellowship. R.A. and R.K. acknowledge the support by the National Science Foundation (NSF) under award number 2004791.

**Conflict of Interest**

The authors declare no conflict of interest. HORIBA is the manufacturer of the AFM equipment used in this study. Collaboration with industry and academia is a part of the job responsibility for M.S. and A.K.

**Data Availability Statement**

The data that support the findings of this study are available from the corresponding author upon reasonable request.

Received: ((will be filled in by the editorial staff))

Revised: ((will be filled in by the editorial staff))

Published online: ((will be filled in by the editorial staff))

References

[1]  C. Battaglia, A. Cuevas, S. De Wolf, *Energy Environ. Sci.* **2016**, *9*, 1552.

[2]  B. M. Kayes, H. Nie, R. Twist, S. G. Spruytte, F. Reinhardt, I. C. Kizilyalli, G. S. Higashi, in *2011 37th IEEE Photovoltaic Specialists Conference*, IEEE, Seattle, WA, USA, **2011**, pp. 000004–000008.

[3]  Z. Chen, B. Turedi, A. Y. Alsalloum, C. Yang, X. Zheng, I. Gereige, A. AlSaggaf, O. F. Mohammed, O. M. Bakr, *ACS Energy Lett.* **2019**, *4*, 1258.

[4]  M. I. Saidaminov, V. Adinolfi, R. Comin, A. L. Abdelhady, W. Peng, I. Dursun, M. Yuan, S. Hoogland, E. H. Sargent, O. M. Bakr, *Nat Commun* **2015**, *6*, 8724.

[5]  Y.-Y. Sun, M. L. Agiorgousis, P. Zhang, S. Zhang, *Nano Lett.* **2015**, *15*, 581.

[6]  W. Meng, B. Saparov, F. Hong, J. Wang, D. B. Mitzi, Y. Yan, *Chem. Mater.* **2016**, *28*, 821.

[7]  S. Niu, H. Huyan, Y. Liu, M. Yeung, K. Ye, L. Blankemeier, T. Orvis, D. Sarkar, D. J. Singh, R. Kapadia, J. Ravichandran, *Adv. Mater.* **2017**, *29*, 1604733.

[8]  S. Perera, H. Hui, C. Zhao, H. Xue, F. Sun, C. Deng, N. Gross, C. Milleville, X. Xu, D. F. Watson, B. Weinstein, Y.-Y. Sun, S. Zhang, H. Zeng, *Nano Energy* **2016**, *22*, 129.

[9]  S. F. Hoefler, G. Trimmel, T. Rath, *Monatsh Chem* **2017**, *148*, 795.

[10] S. Niu, J. Milam-Guerrero, Y. Zhou, K. Ye, B. Zhao, B. C. Melot, J. Ravichandran, *J. Mater. Res.* **2018**, *33*, 4135.

[11] X. Wei, H. Hui, C. Zhao, C. Deng, M. Han, Z. Yu, A. Sheng, P. Roy, A. Chen, J. Lin, D. F. Watson, Y.-Y. Sun, T. Thomay, S. Yang, Q. Jia, S. Zhang, H. Zeng, *Nano Energy* **2020**, *68*, 104317.

[12] C. Comparotto, A. Davydova, T. Ericson, L. Riekehr, M. V. Moro, T. Kubart, J. Scragg, *ACS Appl. Energy Mater.* **2020**, *3*, 2762.

[13] Z. Yu, X. Wei, Y. Zheng, H. Hui, M. Bian, S. Dhole, J.-H. Seo, Y.-Y. Sun, Q. Jia, S. Zhang, S. Yang, H. Zeng, *Nano Energy* **2021**, *85*, 105959.

[14] C. Comparotto, P. Ström, O. Donzel-Gargand, T. Kubart, J. J. S. Scragg, *ACS Appl. Energy Mater.* **2022**, *5*, 6335.

[15] I. Sadeghi, K. Ye, M. Xu, Y. Li, J. M. LeBeau, R. Jaramillo, *Adv Funct Materials* **2021**, *31*, 2105563.

[16] M. Surendran, H. Chen, B. Zhao, A. S. Thind, S. Singh, T. Orvis, H. Zhao, J.-K. Han, H. Htoon, M. Kawasaki, R. Mishra, J. Ravichandran, *Chem. Mater.* **2021**, *33*, 7457.

[17] V. K. Ravi, S. H. Yu, P. K. Rajput, C. Nayak, D. Bhattacharyya, D. S. Chung, A. Nag, *Nanoscale* **2021**, *13*, 1616.

[18] A. A. Pradhan, M. C. Uible, S. Agarwal, J. W. Turnley, S. Khandelwal, J. M. Peterson, D. D. Blach, R. N. Swope, L. Huang, S. C. Bart, R. Agrawal, *Angew Chem Int Ed* **2023**, *62*, e202301049.

[19] S. Dhole, X. Wei, H. Hui, P. Roy, Z. Corey, Y. Wang, W. Nie, A. Chen, H. Zeng, Q. Jia, *Photonics* **2023**, *10*, 366.

[20] S. Niu, B. Zhao, K. Ye, E. Bianco, J. Zhou, M. E. McConney, C. Settens, R. Haiges, R. Jaramillo, J. Ravichandran, *J. Mater. Res.* **2019**, *34*, 3819.

[21] K. Ye, B. Zhao, B. T. Diroll, J. Ravichandran, R. Jaramillo, *Faraday Discuss.* **2022**, 10.1039.D2FD00047D.

[22] W. Melitz, J. Shen, A. C. Kummel, S. Lee, *Surface Science Reports* **2011**, *66*, 1.

[23] W. N. Hansen, G. J. Hansen, *Surface Science* **2001**, *481*, 172.

[24] W. M. Haynes, *CRC Handbook of Chemistry and Physics, 95th Edition*, CRC Press, Hoboken, **2014**.

[25] Z. Zhang, J. T. Yates, *Chem. Rev.* **2012**, *112*, 5520.

[26] N. Gross, Y.-Y. Sun, S. Perera, H. Hui, X. Wei, S. Zhang, H. Zeng, B. A. Weinstein, *Phys. Rev. Applied* **2017**, *8*, 044014.

[27] S. Sharma, Z. Ward, K. Bhimani, K. Li, A. Lakhnot, R. Jain, S.-F. Shi, H. Terrones, N. Koratkar, *ACS Appl. Electron. Mater.* **2021**, *3*, 3306.

[28] G03 Committee, *Tables for Reference Solar Spectral Irradiances: Direct Normal and Hemispherical on 37 Tilted Surface*, ASTM International, **n.d.**

[29] O. E. Semonin, G. A. Elbaz, D. B. Straus, T. D. Hull, D. W. Paley, A. M. Van Der Zande, J. C. Hone, I. Kymissis, C. R. Kagan, X. Roy, J. S. Owen, *J. Phys. Chem. Lett.* **2016**, *7*, 3510.

[30] D. Kang, J. L. Young, H. Lim, W. E. Klein, H. Chen, Y. Xi, B. Gai, T. G. Deutsch, J. Yoon, *Nat Energy* **2017**, *2*, 17043.

Opto-electronic properties of BaZrS$_3$ are measured on single crystals. The illuminated surface shows positive non-contact surface photovoltage, typically observed on *p*-type semiconductors. However, the photocurrent of BaZrS$_3$ is limited by defect formation during processing. Crystal cleavage avoids the extended defects beneath the surface and improves performance. There is room for further optimization by controlling the defects in BaZrS$_3$.

Boyang Zhao, Huandong Chen, Ragib Ashan, Fei Hou, Eric R Hoglund, Shantanu Singh, Huan Zhao, Han Htoon, Andrey Krayev, Maruda Shanmugasundaram, Patrick E Hopkins, Jan Seidel, Rehan Kapadia, and Jayakanth Ravichandran*

**Photoconductive Effects in Single Crystals of BaZrS$_3$**

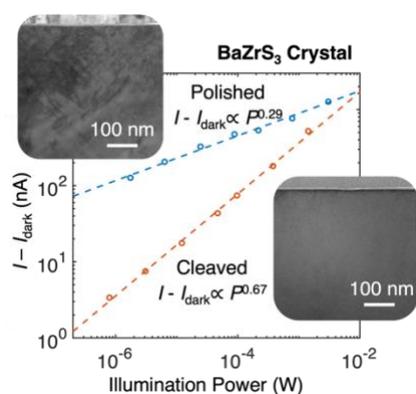

ToC figure ((Please choose one size: 55 mm broad × 50 mm high **or** 110 mm broad × 20 mm high. Please do not use any other dimensions))

# Supporting Information

**Photoconductive Effects in Single Crystals of BaZrS$_3$**

*Boyang Zhao, Huandong Chen, Ragib Ashan, Fei Hou, Eric R Hoglund, Shantanu Singh, Huan Zhao, Han Htoon, Andrey Krayev, Maruda Shanmugasundaram, Patrick E Hopkins, Jan Seidel, Rehan Kapadia, and Jayakanth Ravichandran*

*Boyang Zhao, Huandong Chen, Regib Ashan, Fei Hou, Shantanu Singh, Eric R Hoglund, Huan Zhao, Han Htoon, Andrey Krayev, Maruda Shanmugasundaram, Patrick E Hopkins, Jan Seidel, Rehan Kapadia, and Jayakanth Ravichandran*

Boyang Zhao, Huandong Chen, Shantanu Singh, Jayakanth Ravichandran
Mork Family Department of Chemical Engineering and Materials Science, University of Southern California, Los Angeles, California, 90089, USA
E-mail: j.ravichandran@usc.edu

Regib Ashan, Rehan Kapadia, J. Ravichandran
Ming Hsieh Department of Electrical Engineering, University of Southern California, Los Angeles, California, 90089, USA

Fei Hou, Jan Seidel
School of Materials Science and Engineering, University of New South Wales, Sydney, NSW, Australia

Eric R Hoglund, Patrick E Hopkins
Department of Mechanical and Aerospace Engineering, University of Virginia, Charlottesville, Virginia, 22807, USA
Department of Materials Science and Engineering, University of Virginia, Charlottesville, Virginia, 22904, USA

Huan Zhao, Han Htoon
Center for Integrated Nanotechnologies (CINT), Materials Physics and Applications Division, Los Alamos National Laboratory, Carlsbad, New Mexico, 88220, USA

Andrey Krayev,

HORIBA Instruments Inc., Novato CA, 94949, USA

Maruda Shanmugasundaram

HORIBA Instruments Inc., Piscataway NJ, 08854, USA

Eric R Hoglund

Center for Nanophase Materials Sciences, Oak Ridge National Laboratory, Oak Ridge, Tennessee, 37830, USA

Patrick E Hopkins

Department of Physics, University of Virginia, Charlottesville, Virginia, 22904, USA

Jayakanth Ravichandran

Core Center of Excellence in Nano Imaging, University of Southern California, Los Angeles, California, 90089, USA

## I. Time-Resoled Photoluminescence (TRPL)

TRPL of BaZrS$_3$ crystals can be used to study the relaxation time of photo-excited carriers. Freshly grown BaZrS$_3$ crystals were measured in comparison with single crystals reported ealier[1], as shown in **Figure S1**. Bi-exponential fit is carried out to extract the fast and slow relaxation time. A slower cooling rate of 1℃/h brings the relaxation time ($\tau_1 = 1.069 \pm 0.009$ ns; $\tau_2 = 8.00 \pm 0.15$ ns) of photo-excited states in BaZrS$_3$ single crystals shorter than the crystals grown with a slower cooling rate of 6 ℃/h[1] ($\tau_1 = 0.675 \pm 0.005$ ns; $\tau_2 = 3.97 \pm 0.04$ ns).

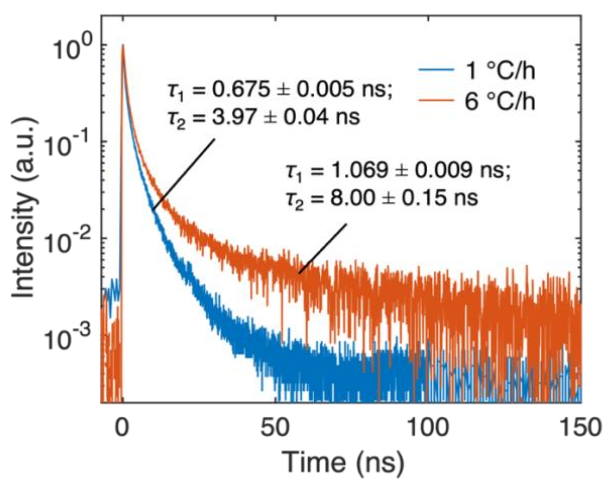

**Figure S1**. TRPL on BaZrS$_3$ crystals grown at 1 ℃/h and 6 ℃/h[1] cooling rates, reproduced with permission. [1] Copyright © 2022 The Royal Society of Chemistry.

## II. Contact Geometry

Electrical contacts collect the photo-excited carriers from the active region. As the mean-free path of photoexcited carriers is predicted to be ~ 5 µm by TRPL, two terminal channels > 10 µm won't be efficiently collecting photoexcited carriers from intrinsic BaZrS$_3$. Electrical contacts could thus be fabricated to a 10 µm channel length with maximized active region, as the finger geometry electrodes [**Figure S2a**] on epitaxial BaZrS$_3$[2].

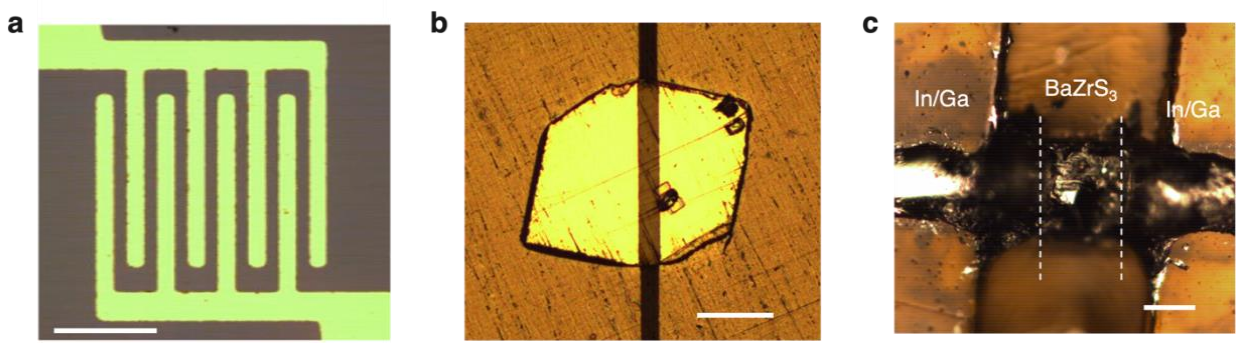

**Figure S2**. Contact geometry of (a) epitaxial BaZrS$_3$[2], reproduced with permission.[2] Copyright © 2021 American Chemical Society; (b) polished BaZrS$_3$; and (c) cleaved BaZrS$_3$. Scale bar 100 μm.

However, polished BaZrS$_3$ crystals have a roughness of a few nm, worse than epitaxial films. With a higher risk of disconnected electrical contacts, we chose to fabricate large metal pads that cover most of the crystalline surface [**Figure S2b**] for a stable and reproducible connection. Although the channel length can be optimized the active region is much smaller than the finger geometry. Conversely, cleaved BaZrS$_3$ has uneven cleavage surfaces that do not support metal deposition. Only the non-ideal two terminal In/Ga electrodes that are manually painted to ~150 μm [**Figure S2c**] junction are tested. The long channel length and randomized contact region are far from an idealized contact geometry.

### III. Kelvin Probe Microscopy (KPFM)

Electrical contact to the back side of BaZrS$_3$ crystals is necessary for surface potential measurements. KPFM detects the surface work function by finding the equilibrium potential between the crystal surface and the AFM probe. We first measure the work function of as-grown BaZrS$_3$, shown in **Figure S3**. On top of flat surfaces of BaZrS$_3$, *CPD* is measured as the DC voltage applied between the crystal and AFM probe to achieve non-contact electron potential equilibrium. The same measurement is carried out on HOPG to get the reference *CPD* of the HOPG work function, which is subtracted to get the work function of BaZrS$_3$.

**Table S1** lists the work function of as-grown BaZrS3 under dark and different laser illumination. Laser power is not quantified. All 450, 532, and 600 nm have higher energy than the absorption bandgap of BaZrS3 and thus excite electron-hole pairs only near the surface of the crystal. The work function changes are listed in **Table S1**.

**Table S1**. The work function of as-grown BaZrS$_3$ single crystal under dark and laser excitation.

| BaZrS$_3$ work function (eV) | | | |
|---|---|---|---|
| no laser | 532nm | 600nm | 450nm |
| 4.44 | 4.26 | 4.29 | 4.31 |

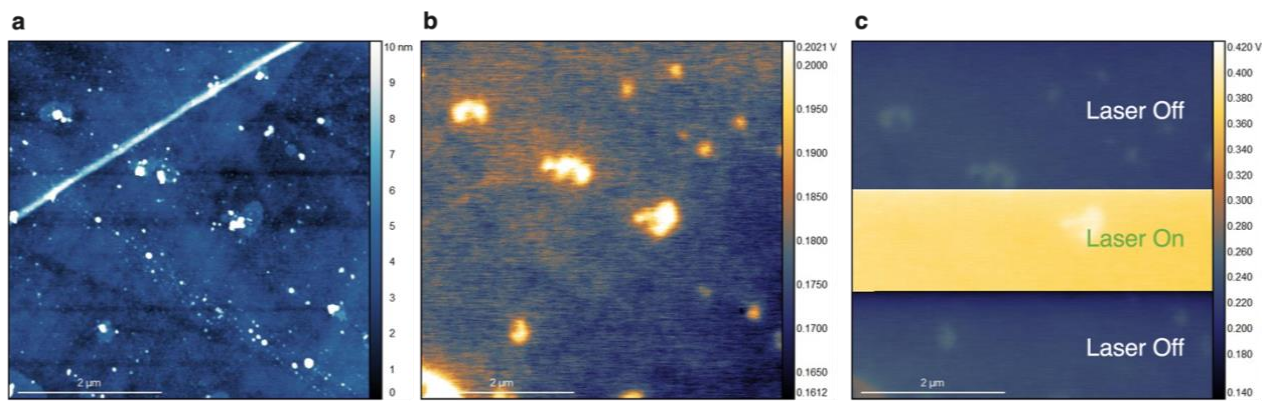

**Figure S3**. KPFM study on as-grown BaZrS₃. (a) Topography of an as-grown BaZrS₃ crystal. (b) *CPD* map of the same region with KPFM. (c) CPD increases under 532 nm illumination.

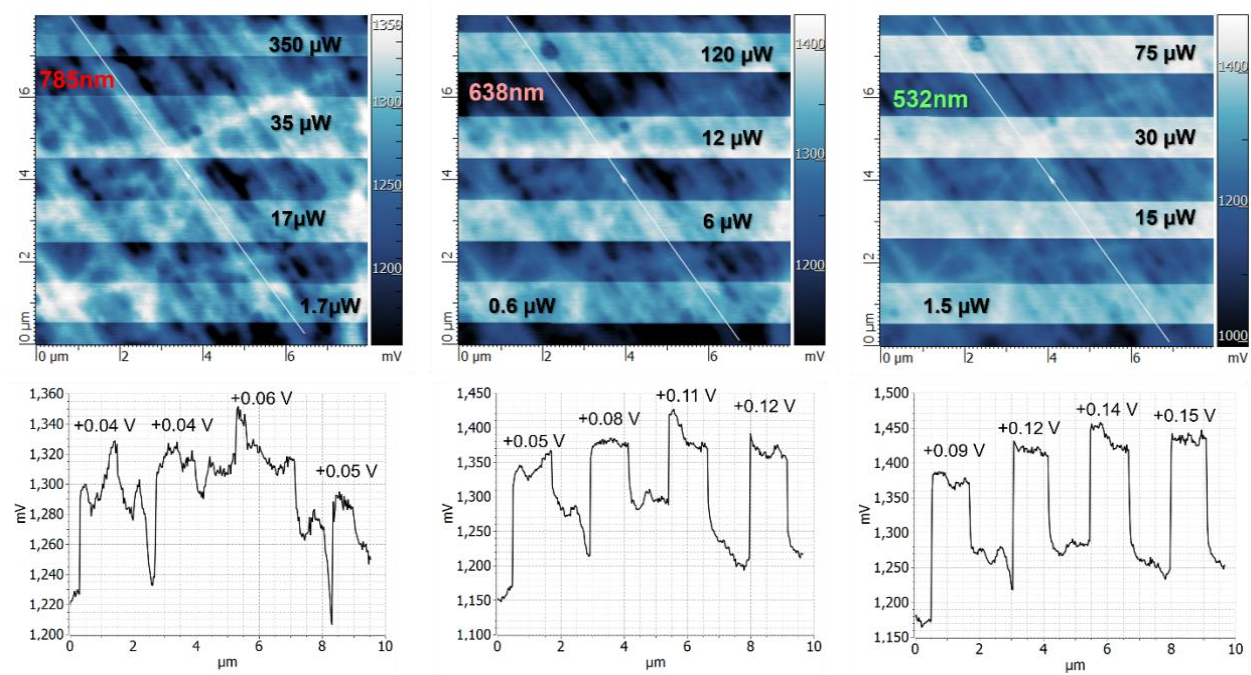

**Figure S4**. KPFM study on polished BaZrS₃ under illumination. (a) Topography of an as-grown BaZrS₃ crystal. (b) *CPD* map of the same region with KPFM.

Quantified laser power of 532 nm, 638 nm, and 785 nm excitation is then tested on KPFM scans of polished BaZrS₃, shown in **Figure S4**. We can observe that 785 nm only slightly increases the *CPD*, whose quantitative accuracy is bad due to being comparable to surface tomography noise. This is because 785 nm is lower energy than the absorption band edge of BaZrS₃ and most light penetrates deeper into BaZrS₃ until being absorbed. 638 nm and 532 nm induce obvious surface potential change, which is further extracted as SPV. Higher power density induces larger *CPD*.

## IV. Scanning Photocurrent Microscopy

Scanning photocurrent microscopy requires a fine laser spot size much smaller than the channel length to gauge the spatial photocurrent profile. We carried out a laser spot size calibration using the intensity of the 521 cm⁻¹ Raman peak of a diced Si wafer. Knife-edge[3] Raman intensity across the edge of a diced Si wafer is mapped at a 0.2 μm step size. The calibration result is shown in **Figure S5**. The laser spot, highly concentrated around ~0.8 μm diameter in the center, is ~ 1.4 μm in diameter.

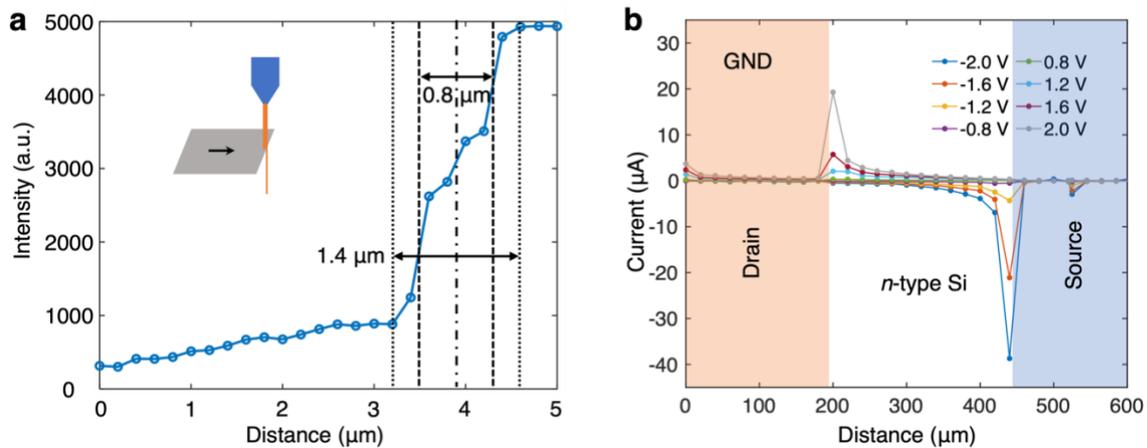

**Figure S5**. SPCM calibration and validation. (a) Laser spot size calibration *via* knife edge 521 cm⁻¹ Raman intensity map across the edge of Si wafer. Raman intensity was detected only when the laser spot was on Si. (b) SPCM of *n*-doped Si wafer. The peak photocurrent is detected near the drain at +V$_{source}$ and near the source for -V$_{source}$, expected for *n*-type degenerately doped semiconductors.

SPCM was first validified on degenerately doped *n*-doped Si (MTI, 1-10 Ω·cm), between two silver paint contacts. **Figure S5b** shows the photocurrent profile when V$_{source}$ ranges from -2 V to +2V. Shaded regions are covered under electrodes while white regions are active channel length. The peak photocurrent is detected next to the drain (+V$_{source}$) or source (-V$_{source}$), which exponentially decays towards the opposite electrode. Such decay corresponds to the minority carrier drift (V$_{sd}$ dependent) or diffusion (V$_{sd}$ independent) length to the drain (+V$_{source}$) or source (-V$_{source}$)[4].

## V. Scanning Tunneling Electron Microscopy (STEM):

The nature of polish-induced defects in BaZrS₃ is analyzed by STEM. Z-contrast imaging [**Figure S6a**] appears uniform while diffraction contrast [**Figure S6b**] shows planner boundaries that are ~45° from 010 orientation. Such boundaries are populated ~300 nm (circled by yellow solid lines) from the surface and turned into darker boundary walls (circled by purple solid lines) deeper into BaZrS₃ as shown in **Figure S6c**. Selected area diffraction (SEAD) is then carried out for both regions. [101]-zone axis electron diffractions of BaZrS₃ are observed but with split diffraction peaks [**Figure S6d** and **Figure S6e**]. This is a sign of small-angle grain boundaries. Moreover, near the polished surface of BaZrS₃, additional diffraction spots are observed that are either broken symmetry of deformed regions or a twined domain of another orientation (i.e. [010]-zone axis). Dislocations, stacking faults and grain boundaries are mechanically formed right beneath the surface.

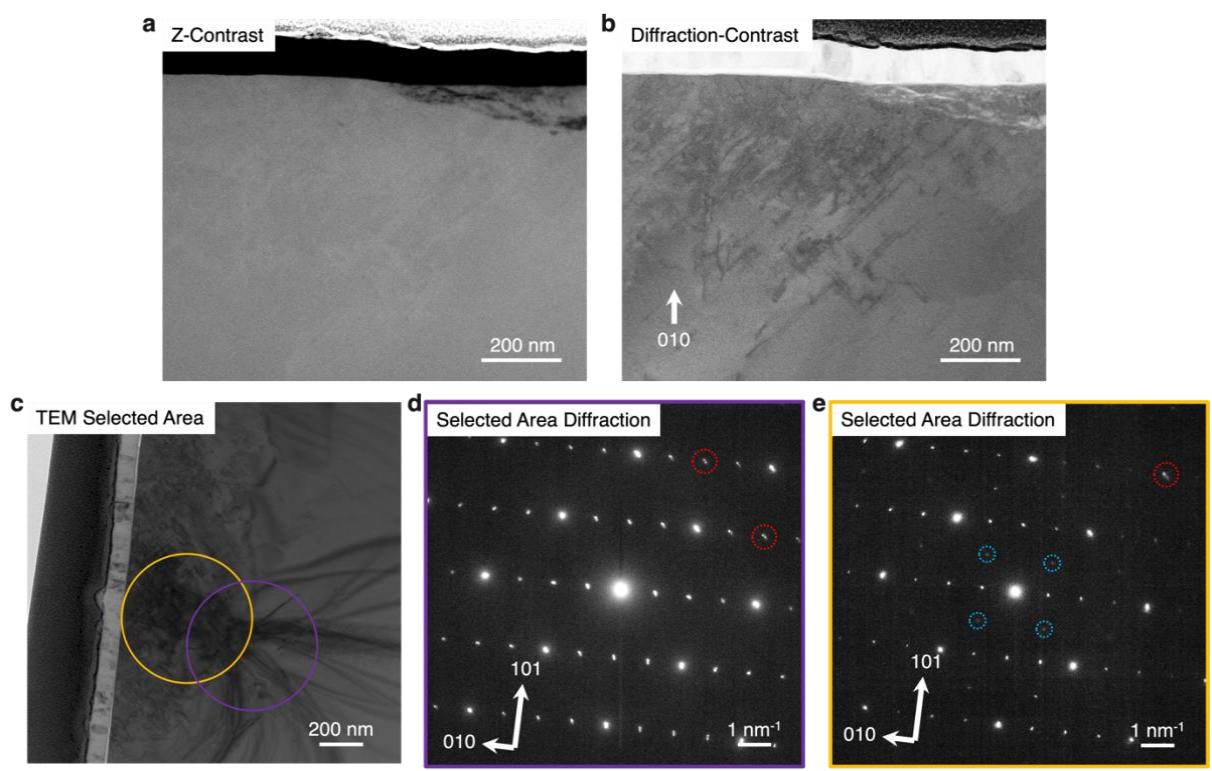

**Figure S6**. STEM of defects in polished BaZrS₃. (a) Z-contrast and (b) diffraction contrast STEM image of the polished BaZrS₃. Although still chemically uniform, polishing caused domain boundaries to form. (c) Defect geometries of polished BaZrS₃. Selected area diffraction (SAED) is carried out for regions circled by yellow and purple solid lines. (d) SEAD of the purple region. Peaks are split, highlighted by red dotted circles. This is a sign of small-angle grain boundaries. (e) SEAD of the yellow region. Aside from the small-angle boundaries, additional reflections are circled by blue dashed lines.

References:

[1]   K. Ye, B. Zhao, B. T. Diroll, J. Ravichandran, R. Jaramillo, *Faraday Discuss.* **2022**, 10.1039.D2FD00047D.

[2]   M. Surendran, H. Chen, B. Zhao, A. S. Thind, S. Singh, T. Orvis, H. Zhao, J.-K. Han, H. Htoon, M. Kawasaki, R. Mishra, J. Ravichandran, *Chem. Mater.* **2021**, *33*, 7457.

[3]   J. M. Khosrofian, B. A. Garetz, *Appl. Opt.* **1983**, *22*, 3406.

[4]   O. E. Semonin, G. A. Elbaz, D. B. Straus, T. D. Hull, D. W. Paley, A. M. Van Der Zande, J. C. Hone, I. Kymissis, C. R. Kagan, X. Roy, J. S. Owen, *J. Phys. Chem. Lett.* **2016**, *7*, 3510.